\documentclass[%
 reprint,
superscriptaddress,
 amsmath,amssymb,
 aps,
nobalancelastpage
]{revtex4-1}

\usepackage[]{graphicx}% Include figure files
\usepackage{dcolumn}% Align table columns on decimal point
\usepackage{bm}% bold math
\usepackage[retainorgcmds]{IEEEtrantools}
\usepackage{times,txfonts}
\usepackage[colorlinks=true,linkcolor=blue,urlcolor=blue,citecolor=blue,pdfusetitle]{hyperref}
\usepackage{mathtools}
\usepackage[%showframe,%Uncomment any one of the following lines to test 
scale=0.9, marginratio={1:1, 2:3}, ignoreall,% default settings
]{geometry}

\DeclareMathOperator{\tr}{tr}
\DeclareMathOperator{\Tr}{Tr}		%Define o operador traço.

\usepackage[version=4]{mhchem}			%formulas quimicas
\usepackage{hyperref}
\newcommand{\subfigimg}[3][,]{%
  \setbox1=\hbox{\includegraphics[#1]{#3}}% Store image in box
  \leavevmode\rlap{\usebox1}% Print image
  \rlap{\hspace*{25pt}\raisebox{\dimexpr\ht1-1.55\baselineskip}{#2}}% Print label
  \phantom{\usebox1}% Insert appropriate spcing
}

\begin{document}

%\preprint{APS/123-QED}

\title{Quantum coherence and criticality in irreversible work}% Force line breaks with \\

\date{\today}
\author{Adalberto D. Varizi}
\affiliation{Instituto de F\'isica da Universidade de S\~ao Paulo,  05314-970 S\~ao Paulo, Brazil}
\affiliation{Departamento de F\'isica, Instituto de Ci\^encias Exatas, Universidade Federal de Minas Gerais, 30123-970, Belo Horizonte, Minas Gerais, Brazil}
\author{Andr\'e P. Vieira}
\affiliation{Instituto de F\'isica da Universidade de S\~ao Paulo,  05314-970 S\~ao Paulo, Brazil}
\author{Cecilia Cormick}
%\email{gtlandi@if.usp.br}
\affiliation{Instituto de F\'isica Enrique Gaviola, CONICET and Universidad Nacional de C\'ordoba, Ciudad Universitaria, X5016LAE C\'ordoba, Argentina}
\author{Raphael C. Drumond}
%\email{gtlandi@if.usp.br}
\affiliation{Departamento de Matem\'atica, Instituto de Ci\^encias Exatas, Universidade Federal de Minas Gerais, 30123-970, Belo Horizonte, Minas Gerais, Brazil}
\author{Gabriel T. Landi}
%\email{gtlandi@if.usp.br}
\affiliation{Instituto de F\'isica da Universidade de S\~ao Paulo,  05314-970 S\~ao Paulo, Brazil}

\begin{abstract}
The irreversible work during a driving protocol constitutes one of the most widely studied measures in non-equilibrium thermodynamics, as it constitutes a proxy for entropy production.  
%When a system undergoes a work protocol it  usually lags behind equilibrium due to the finite-time nature of the drive
% { \color{red}[Is this true for many-body systems? That is, for many-body systems, even if the drive takes infinity time, the system lags behind. Or not?]}. 
%This non-equilibrium lag is proportional to the irreversible work and thus serves as a proxy for entropy production.
In quantum systems, it has been shown that the irreversible work has an additional, genuinely quantum mechanical contribution, due to coherence produced by the driving protocol. 
The goal of this paper is to explore this contribution in systems that undergo a quantum phase transition. 
Substantial effort has been dedicated in recent years to understand the role of quantum criticality in work protocols. However, practically nothing is known about how  coherence contributes to it. 
To shed light on this issue, we study the entropy production in infinitesimal quenches of the one-dimensional XY model.
For quenches in the transverse field, we find that for finite temperatures the contribution from coherence can, in certain cases, account for practically all of the entropy production.
At low temperatures, however, the coherence presents a finite cusp at the critical point, whereas the entropy production diverges logarithmically. 
Alternatively, if the quench is performed in the anisotropy parameter, we find that there are situations where all of the entropy produced is due to quantum coherences. 

\end{abstract}

\maketitle

\section{\label{SecIntro}Introduction}

\begin{figure}
\centering
\includegraphics[width=0.4\textwidth]{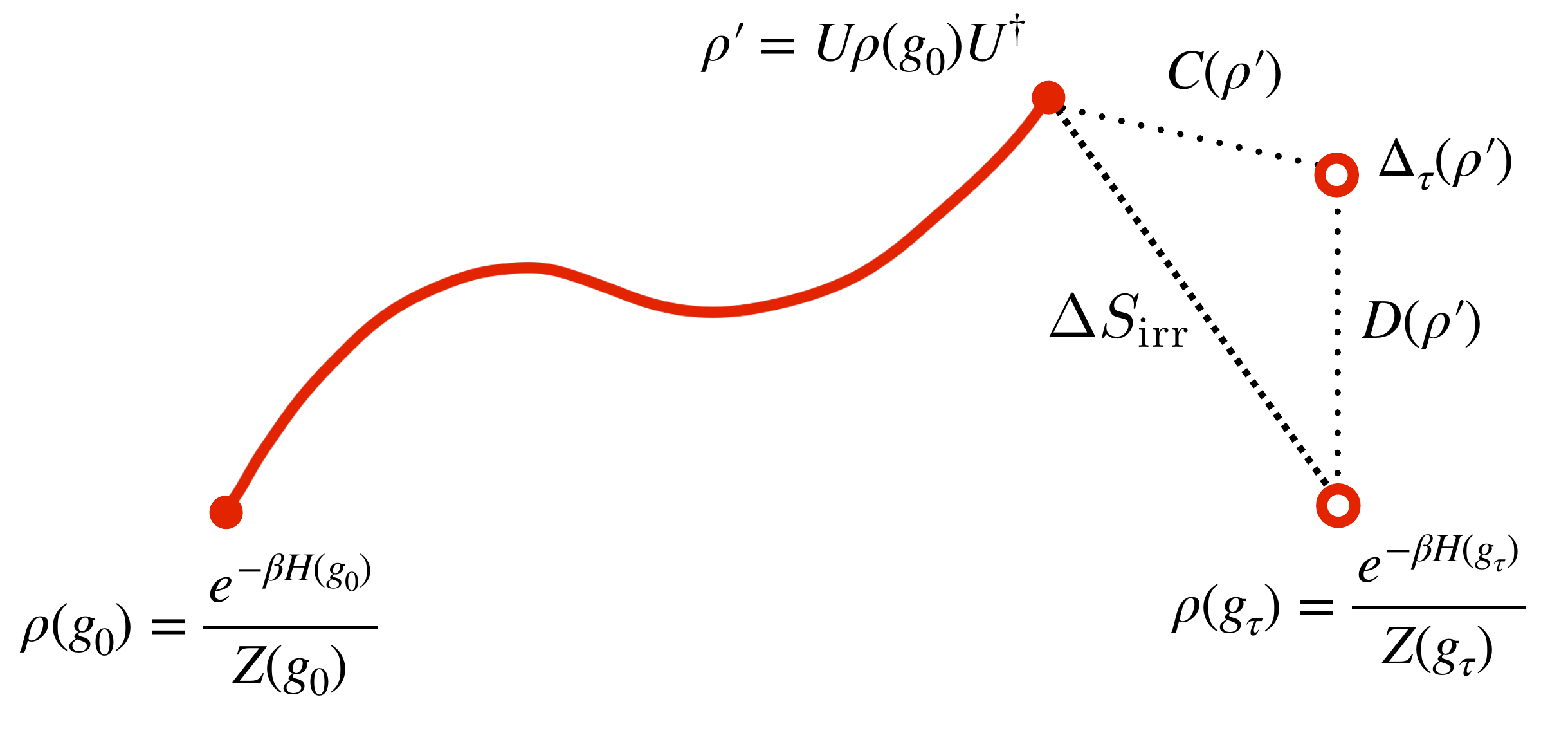}
\caption{\label{fig:drawing}
Irreversible work and entropy production. 
A system with Hamiltonian $H(g)$ is initially prepared in equilibrium at a temperature $\beta = 1/T$, with $g=g_0$.
The system is then pushed out of equilibrium by means of a work protocol $g(t)$, which lasts for a total duration $\tau$. This  generates a unitary $U$ which pushes the system away from equilibrium to a state  $\rho' = U \rho(g_0) U^\dagger$.
The non-equilibrium nature of the process can be quantified by the irreversible work $W_\text{irr}$ [Eq.~(\ref{Wirr})] or, what is equivalent, the entropy production/non-equilibrium lag $\Delta S_\text{irr}$ in Eq.~(\ref{DeltaSirr}). 
This quantity, however, can be split as in Eq.~(\ref{DeltaS_splitting}) into a contribution $D$ [Eq.~(\ref{D})] related to changes in the population and a contribution $C$ related to quantum coherence [Eq.~(\ref{C})].
}
\end{figure}

Driving a system out of equilibrium is always accompanied by a finite production of entropy. 
The typical scenario is that shown in Fig.~\ref{fig:drawing}. 
A system with Hamiltonian $H(g)$, depending on an externally tunable parameter $g$, is initially prepared in thermal equilibrium at a temperature $T$, so that its density matrix is given by $\rho(g_0) = e^{-\beta H(g_0)}/Z(g_0)$, where $\beta = 1/T$ and $Z(g_0)$ is the partition function. 
At $t=0$ the system is  driven out of equilibrium by changing $g$ according to some work protocol $g(t)$ that lasts for a total time $\tau$. 
If the dynamics can be considered unitary, the state of the system after the drive will be
\begin{equation}\label{rho_tau}
\rho' = U \rho(g_0) U^\dagger, 
\end{equation}
where $U = \mathcal{T} e^{-i \int_0^\tau H(g(t)) dt}$ is the time-evolution operator (with $\mathcal{T}$ standing for the time-ordering operator). 
This state is generally far from the corresponding equilibrium state $\rho(g_\tau)$; the difference between them can be quantified by the irreversible work \cite{Jarzynski1997,Kurchan1998,Talkner2007}
\begin{equation}\label{Wirr}
W_\text{irr} = \langle W \rangle - \Delta F, 
\end{equation} 
where $\langle W \rangle = \tr\big\{ H(g_\tau) \rho' - H(g_0) \rho(g_0)\big\}$ is the average work performed in the process and $\Delta F = - T \ln Z(g_\tau)/Z(g_0)$ is the change in equilibrium free energy. 
Eq.~(\ref{Wirr}) can also be written solely in terms of information theoretic quantities (called the \emph{non-equilibrium lag}), as  
\cite{Kawai2007,Vaikuntanathan2009,Parrondo2009,Deffner2010,Batalhao2015}
\begin{equation}\label{DeltaSirr}
\Delta S_\text{irr} = \beta W_\text{irr} =  S(\rho' || \rho(g_\tau)),
\end{equation}
where $S(\rho||\sigma) = \tr(\rho \ln \rho - \rho \ln \sigma)$ is the quantum relative entropy. 
It thus measures the entropic distance between the final state $\rho'$ and the associated equilibrium state $\rho(g_\tau)$ that the system does not tend to since the process is out of equilibrium (Fig.~\ref{fig:drawing}). 
Since $S(\rho' || \rho(g_\tau)) \geq 0$ by construction, this shows quite clearly why $\Delta S_\text{irr}$ or $W_\text{irr}$ can be used to quantify the non-equilibrium nature of the process \cite{Fermi1956,Jarzynski1997,Talkner2007}.

Strictly speaking, since the dynamics is unitary, no entropy is produced in the map~(\ref{rho_tau}). 
The non-equilibrium lag~(\ref{DeltaSirr}) is nonetheless a proxy for the entropy production. 
The reason is that, if after the protocol the system is once again coupled  to a bath, it will relax from $\rho'$ to $\rho(g_\tau)$, a process whose entropy production is precisely $\Delta S_\text{irr}$ in Eq.~(\ref{DeltaSirr}) \cite{Spohn1978,Breuer2003,Santos2019}. 
For this reason, even though the process~(\ref{rho_tau}) is unitary, one commonly associates $\Delta S_\text{irr}$ with its entropy production.

This typical work-protocol scenario has been the subject of countless studies, both theoretical~\cite{Jarzynski1997,Jarzynski1997a,Derrida1998,Crooks1998,Kurchan1998,Lebowitz1999,Jarzynski1999a,Maes1999,Jarzynski2001a,Crooks2000,Jarzynski2000,Mukamel2003,Andrieux2004,Monnai2005,Teifel2007,Talkner2007,Kawai2007,Crooks2008,Gelin2008,Jarzynskia2008,Talkner2009,Vaikuntanathan2009,Parrondo2009,Deffner2010,Teifel2011,Mazzola2013a,Dorner2013,Talkner2013a,Sivak2012a,Hoppenau2013,Watanabe2014,Roncaglia2014a,Plastina2014,Skrzypczyk2014,Halpern2015,Solinas2015,Funo2015,Alhambra2016,Alhambra2016a,Talkner2016,Jin2016a,Chenu2017a,Solinas2017,Bartolotta2017,Park2017,Perarnau-Llobet2017,Sampaio2017,Elouard2017a,DiStefano2017,Wei2017a,Elouard2017,Lostaglio2018,Guarnieri2018,Manzano2017a,Francica2019,DeChiara2018a,Fusco2014a,Francica2016a,Zhong2015,Apollaro2014,Solinas2015,DelCampo2014,Brunelli2014,Allahverdyan2005,Crooks2007,Talkner2008,Deffner2008,Dorosz2008a,Dorner2012,Ryabov2013,Carlisle,Roncaglia2014a,Perarnau-Llobet2014,Sindona2014,Chiara2015,Arrais2018,obejko2017,Bayocboc2015} as well as experimental~\cite{Liphardt2002,Douarche2005,Collin2005,Speck2007,Saira2012,Koski2013,Batalhao2014,An2014,Batalhao2015,Talarico2016,Zhang2018a,Smith2017}
However, although Eq.~(\ref{DeltaSirr}) is formulated for quantum systems, many aspects of it are often classical. 
The issue of what are the genuinely quantum features of such a process, despite still being the subject of debate, is ultimately related to the notion of quantum coherence. 
The thermodynamic processes involved in the map~(\ref{rho_tau}) highlight the energy basis as a preferred basis (in the sense of~\cite{Zurek1981}). 
Coherence in the energy basis therefore represents the key feature distinguishing classical and quantum processes~\cite{Lostaglio2015,Santos2019}.
As the system is driven by the work protocol $g(t)$, the eigenbases of $H(g(t))$ at different times are not necessarily compatible, a feature which has no classical counterpart~\cite{Fusco2014a}.

Several results have recently appeared, which highlight the non-trivial role of coherence in irreversible thermodynamics. 
For instance, Ref.~\cite{Miller2019,Scandi2019} considered quasi-static drives and showed how the standard fluctuation-dissipation theorem is modified to include a term related to $[H(g(t)), dH(g(t))/dt]$, thus reflecting the basis incompatibility during the drive. 
In Ref.~\cite{Santos2019} some of us have shown that during relaxation to equilibrium, the presence of initial coherences contributes an additional term to the entropy production. 
A similar effect also occurs for unitary drives and the non-equilibrium lag, as shown in~\cite{Francica2019}.
In this case,  Eq.~\eqref{DeltaSirr} may quite generally be decomposed as 
\begin{equation}\label{DeltaS_splitting}
 \Delta S_{\text{irr}}=D(\rho') +C(\rho').
\end{equation}
The first term quantifies the contribution   from  changes in the population of the system and reads
\begin{equation}\label{D}
D(\rho') = S(\Delta_{\tau}[\rho']||\rho(g_{\tau})),
\end{equation}
where  $\Delta_{\tau}[\rho']$ is the completely dephased state, obtained from $\rho'$ by eliminating its off-diagonal terms in the eigenbasis of $H(g_{\tau})$.
The second term in Eq.~(\ref{DeltaS_splitting}), on the other hand, is the \emph{relative entropy of coherence},  given by 
\begin{equation}\label{C}
C(\rho')=S(\rho'||\Delta_{\tau}[\rho'])=S(\Delta_{\tau}[\rho'])-S(\rho').
\end{equation}
It therefore quantifies the difference between $\rho'$ and the dephased state $\Delta_\tau[\rho']$. 
This term therefore measures the contribution to the non-equilibrium lag stemming solely from the quantum coherences generated by the  driving protocol.
Since both terms are individually non-negative by construction, this shows how coherence increases the entropy produced in the process. 

In this work we will be interested in the relative contributions of the two terms in Eq.~(\ref{DeltaS_splitting}) in the specific case of  quantum critical systems undergoing infinitesimal quenches.
That is, when the control parameter changes instantaneously from $g_0 \to g_\tau = g_0 + \delta g$, where $\delta g \ll g_0$. 
As shown in Refs.~\cite{Gambassi1106,Dorner2012,PhysRevX4031029}, the non-equilibrium lag simplifies considerably in this case, since one removes the generally complicated dependence on the exact form of the work protocol $g_\tau$.  
Notwithstanding, the problem still retains several interesting features, particularly for quantum critical systems, as beautifully shown in Refs.~\cite{Dorner2012,Mascarenhas1307}.
This has led to a large number of studies on the critical properties of $\Delta S_\text{irr}$ in several models~\cite{PhysRevE92022108,PhysRevA95063615,Pagnelli,PhysRevE98022107,PhysRevB93201106,Bayocboc2015,PhysRevE97052148,Nigro,PhysRevA99043603}. 
A proposal to measure it experimentally in ultra-cold atoms was also given in~\cite{Villa2018}. 

None of the studies above, however, dealt with the relative contribution from populations and coherences [Eq.~(\ref{DeltaS_splitting})]. 
How relevant is $C(\rho')$ therefore remains unknown, even for the simplest critical models.  
It is the goal of this paper to fill in this gap and carry out a detailed study of the contribution from quantum coherence to the non-equilibrium lag in critical infinitesimal quenches. 
To accomplish this, we focus on the  one-dimensional  XY spin chain \cite{Lieb1961}.
The advantage of this model is that by tuning the anisotropy parameter one may tune the relative contribution of $C(\rho')$ when going from the XX to the transverse field Ising model. 
We show that for intermediate and high temperatures, both terms in Eq.~(\ref{DeltaS_splitting}) contribute similarly to $\Delta S_\text{irr}$. 
At low temperatures, on the other hand, $C(\rho')$ becomes sub-dominant. 
And while  $D(\rho')$ diverges logarithmically at the critical point~\cite{Mascarenhas1307,Bayocboc2015}, $C(\rho')$ presents a cusp (i.e., its derivative is discontinuous).

\section{\label{SecModel}Basic setup}
The Hamiltonian of the ferromagnetic $XY$ model may be written as
\begin{equation}\label{EqModel}
 H(g,\,\gamma)=-\sum_{j=1}^{N}\bigg(\frac{1+\gamma}{2}\sigma_j^x\sigma_{j+1}^x+\frac{1-\gamma}{2}\sigma_j^y\sigma_{j+1}^y+g\sigma_j^z\bigg),
\end{equation}
where $\sigma_j^a$ ($a=x,\,y,\,z$) are Pauli spin operators, $N$ is the total number of spins, $\gamma\in[0,\,1]$ is the anisotropy parameter of the spin interaction and $g$ is the applied magnetic field. 
We assume $N$ even, with periodic boundary conditions.
This model presents a paramagnetic phase when $|g|>1$ and a ferromagnetic phase for $|g|<1$, with critical points at $g^*=\pm1$. Special cases occur when one makes $\gamma=0$, to get the XX chain, and $\gamma=1$, to get the Ising model.

The Hamiltonian~(\ref{EqModel}) is diagonalized by introducing  the Jordan-Wigner transformation \cite{qpt}, that maps the spin chain onto an equivalent system of spinless fermions,
\begin{equation}\begin{aligned}
 &\sigma^x_j=(\hat c_j^{\dagger}+\hat c_j)\prod_{i<j}(1-2\hat c_i^{\dagger}\hat c_i),\\
 &\sigma^y_j=\imath(\hat c_j^{\dagger}-\hat c_j)\prod_{i<j}(1-2\hat c_i^{\dagger}\hat c_i),\quad\sigma^z_j=1-2\hat c_j^{\dagger}\hat c_j,
\end{aligned}\end{equation}
where $\hat c_j^{\dagger}$ and $\hat c_j$ are canonical creation and annihilation fermionic operators. After this, one finds that the Hamiltonian \eqref{EqModel} may be broken into two parts belonging to the orthogonal subspaces of positive and negative parity - i.e. subspaces of states with even or odd number of $c$-particles (or up spins), respectively. 
Each part can be independently diagonalized by a Fourier transform followed by a Bogoliubov transformation \cite{Damski}. 
However, they differ only by boundary terms which become negligible in the thermodynamic limit ($N\rightarrow\infty$). 
Hence, all calculations may therefore be performed considering only the positive parity subspace. We therefore consider here that, after diagonalization, we simply have
\begin{equation}\begin{aligned}
  H(g,\,\gamma)=&\sum_{k\in K^+}\epsilon_k(g,\,\gamma)\big(2\eta_k^{\dagger}\eta_k-1\big),
\end{aligned}\end{equation}
where $K^+=\big\{k=\pm(2n+1)\pi/N;\, n=0,\,1,\,2,\,...,N/2-1\big\}$. 
The dispersion relation $\epsilon_k(g,\,\gamma)$ is given by
\begin{subequations}\begin{align}
 &\epsilon_k(g,\,\gamma)=\sqrt{[g-\cos(k)]^2+\gamma^2\sin^2(k)},
\intertext{
and the canonical fermionic operators $\{\eta_k\}$, \emph{which depend on $g$ and $\gamma$}, are given by
}
 &\eta_k=\cos(\theta_k/2)\hat c_k+\sin(\theta_k/2)\hat c_{-k}^{\dagger},
\intertext{
where
}
 &\Big(\sin\theta_k,\,\cos\theta_k\Big)=\bigg(\frac{\gamma\sin(k)}{\epsilon_k(g,\,\gamma)},\,\frac{g-\cos(k)}{\epsilon_k(g,\,\gamma)}\bigg)\label{BogAngles},
 \intertext{
 and
} 
 &\hat c_j=\frac{e^{-\imath\pi/4}}{\sqrt{N}}\sum_{k\in K^+}\hat c_ke^{\imath kj}.\label{FourierTransform}
\end{align}\end{subequations}
For the special case $\gamma=0$, a Bogoliubov transformation is not necessary since the Hamiltonian $H_{\gamma=0}$ becomes  diagonal after the Fourier transformation \eqref{FourierTransform}, and is given by
\begin{equation}\label{XXHamil}
 H_{\gamma=0}(g)=\sum_{k\in K^+}(g-\cos k)\big(2\hat c_k^{\dagger}\hat c_k-1\big).
\end{equation}

Our goal is to compute the entropic quantities appearing in Eqs.~(\ref{D}) and (\ref{C}) for a quantum quench protocol. 
We initially consider the system to have an anisotropy parameter $\gamma_0$, transverse field $g_0$ and to be in equilibrium with a thermal reservoir at inverse temperature $\beta$. 
%We write $\hat{H}(0) = \hat{H}(g_0, \gamma_0)$ and similarly for all other quantities.
The initial state of the spin chain is therefore the thermal state $\rho(0)=e^{-\beta H(0)}/Z(0)$, with $H(0)=H(g_0,\,\gamma_0)$ and partition function $Z(0)=\Tr[e^{-\beta H(0)}]$. Thus can be further decomposed as 
\begin{subequations}\label{Gibbs0}
 \begin{align}
 \rho(0)&=\bigotimes_{\overset{k\in K^+}{k>0}}\rho_{\pm k}^{0},\\
 \rho_{\pm k}^0&=\frac{1}{Z_k^2(0)}\sum_{n_{\pm k}=0}^1e^{2\beta\epsilon_k^0(1-n_k-n_{-k})}\arrowvert n_{-k}n_k\rangle\langle n_{-k}n_k\arrowvert,
 \end{align}\end{subequations}
where  $\arrowvert n_{-k}n_k\rangle$ and $\epsilon_k^0 = \epsilon_k(g_0,\gamma_0)$ are the eigenstates and eigenenergies of  $H(0)$ and $Z_k(0)=2\cosh\big(\beta\epsilon_k^0\big)$. 
The initial von Neumann entropy of this state is thus given by
\begin{equation}\label{EntIni}
S(\rho(0))=\sum_{\overset{k\in K^+}{k>0}}S(\rho_{\pm\,k}^0)=2\sum_{\overset{k\in K^+}{k>0}}\Big[\ln Z_k(0)-\beta\epsilon_k^0\tanh\big(\beta\epsilon_k^0\big)\Big].
\end{equation}

At $t=0$ the system is decoupled from the thermal reservoir and undergoes a sudden quench, where the field is instantaneously changed to $g_\tau$ and/or the anisotropy to $\gamma_\tau$. 
The Hamiltonian therefore changes from $H(g_0,\,\gamma_0)$ to $H(g_{\tau},\,\gamma_\tau)$.
Moreover, since we are considering a sudden quench, the state of the system does not change, so that $\rho' = \rho(0)$. 
However, since in general $[H(g_0,\gamma_0), H(g_\tau, \gamma_\tau)]\neq 0$, the state  $\rho'$ will no longer be diagonal in the eigenbasis of $H(g_\tau, \gamma_\tau)$. 
To express $\rho'$ in the new basis we first note that the  post quench fermionic operators $\{\tilde{\eta}_k\}$ are related to the pre-quench operators $\{\eta_k\}$ according to 
\begin{equation}\label{operatorsmix}
 \tilde{\eta}_k=\eta_k\cos(\Delta_k/2)+\eta_{-k}^{\dagger}\sin(\Delta_k/2),
\end{equation}
where $\Delta_k=\tilde{\theta}_k-\theta_k$ is the difference between the post- and pre-quench Bogoliubov angles \eqref{BogAngles} and can be written as 
\begin{equation}
 \sin\Delta_k=\frac{\sin k}{\epsilon_k^\tau\epsilon_k^0}\big[\gamma_{\tau}(g_0-\cos k)-\gamma_0(g_{\tau}-\cos k)\big],
\end{equation}
with $\epsilon_k^\tau=\epsilon_k(g_{\tau},\,\gamma_\tau)$.
As a consequence the pre- and post-quench eigenstates will be related by 
\begin{IEEEeqnarray}{rCl}
\nonumber
 \arrowvert0_{-k}0_k\rangle&=&\cos(\Delta_k/2)\arrowvert\tilde{0}_{-k}\tilde{0}_k\rangle-\sin(\Delta_k/2)\arrowvert\tilde{1}_{-k}\tilde{1}_k\rangle,\\[1em]
 \label{pre_post_states}
 \arrowvert1_{-k}1_k\rangle&=&\sin(\Delta_k/2)\arrowvert\tilde{0}_{-k}\tilde{0}_k\rangle+\cos(\Delta_k/2)\arrowvert\tilde{1}_{-k}\tilde{1}_k\rangle,\\[1em]
 \arrowvert0_{-k}1_k\rangle&=&\arrowvert\tilde{0}_{-k}\tilde{1}_k\rangle,\qquad%\\[1em]
% &
 \arrowvert1_{-k}0_k\rangle=\arrowvert\tilde{1}_{-k}\tilde{0}_k\rangle.
 \nonumber
\end{IEEEeqnarray}
\begin{widetext}

Using this in Eq.~(\ref{Gibbs0}) we then find
\begin{equation}\label{rho_prime}
\begin{aligned}
\rho'=\bigotimes_{\overset{k\in K^+}{k>0}}\tilde{\rho}_{\pm k},&\\[1em]
\tilde{\rho}_{\pm k}=\frac{1}{Z_k^2(0)}\Bigg\{&\arrowvert\tilde{0}_{-k}\tilde{0}_k\rangle\langle\tilde{0}_{-k}\tilde{0}_k\arrowvert\Big[\cosh\big(2\beta\epsilon_k^0\big)+\sinh\big(2\beta\epsilon_k^0\big)\cos\Delta_k\Big]+\arrowvert\tilde{1}_{-k}\tilde{1}_k\rangle\langle\tilde{1}_{-k}\tilde{1}_k\arrowvert\Big[\cosh\big(2\beta\epsilon_k^0\big)-\sinh\big(2\beta\epsilon_k^0\big)\cos\Delta_k\Big]\\[1em]
+&\arrowvert\tilde{0}_{-k}\tilde{1}_k\rangle\langle\tilde{0}_{-k}\tilde{1}_k\arrowvert+\arrowvert\tilde{1}_{-k}\tilde{0}_k\rangle\langle\tilde{1}_{-k}\tilde{0}_k\arrowvert-\Big(\arrowvert\tilde{0}_{-k}\tilde{0}_k\rangle\langle\tilde{1}_{-k}\tilde{1}_k\arrowvert+\arrowvert\tilde{1}_{-k}\tilde{1}_k\rangle\langle\tilde{0}_{-k}\tilde{0}_k\arrowvert\Big)\sinh(2\beta\epsilon_k^0)\sin(\Delta_k)\Bigg\}.
\end{aligned}\end{equation}
We now use this to compute the relative entropy of coherence in Eq.~(\ref{C}). 
The state  $\Delta_\tau[\rho']$ is obtained by taking only the diagonal entries of Eq.~(\ref{rho_prime}). 
As a consequence, one readily finds that 
\begin{equation}\label{EntDia}\begin{aligned}
 &S(\Delta_{\tau}[\rho'])=\sum_{\overset{k\in K^+}{k>0}}\bigg\{2\ln Z_k(0)-\frac{1}{2}\tanh\big(\beta\epsilon_k^0\big)\cos(\Delta_k)\ln\bigg[\frac{1+\tanh\big(2\beta\epsilon_k^0\big)\cos(\Delta_k)}{1-\tanh\big(2\beta\epsilon_k^0\big)\cos(\Delta_k)}\bigg]-\frac{\cosh\big(2\beta\epsilon_k^0\big)}{4\cosh^2\big(\beta\epsilon_k^0\big)}\ln\Big[1+\sinh^2\big(2\beta\epsilon_k^0\big)\sin^2(\Delta_k)\Big]\bigg\},
\end{aligned}\end{equation}
Eq.~(\ref{C}) then follows from  subtracting~(\ref{EntIni}) from~(\ref{EntDia}). 
We focus on the thermodynamic limit ($N\rightarrow\infty$), where all $k$-sums may be converted into integrals. 
Moreover, we study the relative entropy of coherence \emph{per particle} as $\mathcal{C}(\rho') = C(\rho')/N$. 
In the limit $N\to \infty$ one then finds
\begin{equation}\label{GenCoh}
 \begin{aligned}
 \mathcal{C}(\rho')=\int_0^{\pi}\frac{\mathrm{d}k}{2\pi}\,\Bigg\{&\frac{1}{2}\tanh\big(\beta\epsilon_k^0\big)\Bigg[\ln\bigg[\frac{1+\tanh\big(2\beta\epsilon_k^0\big)}{1-\tanh\big(2\beta\epsilon_k^0\big)}\bigg]-\cos(\Delta_k)\ln\bigg[\frac{1+\tanh\big(2\beta\epsilon_k^0\big)\cos(\Delta_k)}{1-\tanh\big(2\beta\epsilon_k^0\big)\cos(\Delta_k)}\bigg]\Bigg]%\\[1em]
 %
 %&
 -\frac{\cosh\big(2\beta\epsilon_k^0\big)}{4\cosh^2\big(\beta\epsilon_k^0\big)}\ln\Big[1+\sinh^2\big(2\beta\epsilon_k^0\big)\sin^2(\Delta_k)\Big]\Bigg\}.
 \end{aligned}
\end{equation}

A similar calculation was  done for the non-equilibrium lag $ \Delta\mathcal{S}_{\text{irr}} =  \Delta S_{\text{irr}}/N$ in Ref.~\cite{Bayocboc2015}, which found
\begin{equation}\label{GenEnt}\begin{aligned}
 \Delta\mathcal{S}_{\text{irr}}=\int_0^{\pi}\frac{\mathrm{d}k}{2\pi}\,2\bigg\{\ln\bigg[\frac{\cosh\big(\beta\epsilon_k^\tau\big)}{\cosh\big(\beta\epsilon_k^0\big)}\bigg]+\beta\big(\epsilon_k^0-\epsilon_k^\tau\cos(\Delta_k)\big)\tanh\big(\beta\epsilon_k^0\big)\Bigg\}.
\end{aligned}\end{equation}
From~(\ref{GenCoh}) and (\ref{GenEnt}), $D(\rho')$ in Eq.~(\ref{D})  can be readily computed using Eq.~(\ref{DeltaS_splitting}). 
Focusing again on the contribution per particle, $\mathcal{D}(\rho') = D(\rho')/N$, one then finds 
\begin{equation}\label{GenPop}
 \begin{aligned}
 \mathcal{D}(\rho')=\int_0^{\pi}\frac{\mathrm{d}k}{2\pi}\,\Bigg\{&2\ln\bigg[\frac{\cosh\big(\beta\epsilon_k^\tau\big)}{\cosh\big(\beta\epsilon_k^0\big)}\bigg]-\frac{1}{2}\tanh\big(\beta\epsilon_k^0\big)\cos(\Delta_k)\Bigg[\ln\bigg[\frac{1+\tanh\big(2\beta\epsilon_k^{\tau}\big)}{1-\tanh\big(2\beta\epsilon_k^{\tau}\big)}\bigg]-\ln\bigg[\frac{1+\tanh\big(2\beta\epsilon_k^0\big)\cos(\Delta_k)}{1-\tanh\big(2\beta\epsilon_k^0\big)\cos(\Delta_k)}\bigg]\Bigg]\\[0.8em]
 &
 +\frac{\cosh\big(2\beta\epsilon_k^0\big)}{4\cosh^2\big(\beta\epsilon_k^0\big)}\ln\Big[1+\sinh^2\big(2\beta\epsilon_k^0\big)\sin^2(\Delta_k)\Big]\Bigg\}.
 \end{aligned}
\end{equation}
As a sanity check, in the case of an  XX chain ($\gamma_0 = \gamma_\tau= 0$) the quench does not affect the eigenbasis so $\Delta_k = 0$.
Hence, $\mathcal{C}(\rho') = 0$, and all contributions to the non-equilibrium lag stems from the changes in populations.

\end{widetext}

%%%%%%%%%%%%%%%%%%%%%%%%%%%%%%%%%%%%
\section{\label{SecHighLow}High and low temperature limits}

Since these results are somewhat complicated, we now proceed to separately analyze some limiting cases. 
As a consistency check, in all numerical analyses presented in this section, the integral expressions~(\ref{GenCoh})-(\ref{GenPop}) were compared with exact numerics; i.e., obtained from discrete summations over the set $K^+$ [c.f.~Eq.~(\ref{EntDia})] for sufficiently large $N$.

\subsection{\label{SmallBeta}High temperature limit}
For small $\beta$ (high temperatures), the expressions for $\mathcal{C}(\rho')$, $\mathcal{D}(\rho')$ and $\Delta\mathcal{S}_{\text{irr}}$ simplify dramatically to 
\begin{subequations}\begin{align}\label{CSb}
  &\mathcal{C}(\rho')= \beta^2 \int_0^{\pi}\frac{\mathrm{d}k}{2\pi}\,(\epsilon_k^0)^2\sin^2\Delta_k,\\[0.8em]
  \label{DSb}
  &\mathcal{D}(\rho')=\beta^2 \int_0^{\pi}\frac{\mathrm{d}k}{2\pi}\,\big(\epsilon_k^\tau-\epsilon_k^0\cos\Delta_k\big)^2,\\[0.8em]
  \label{IrrSb}
  &\Delta\mathcal{S}_{\text{irr}}= \beta^2 \int_0^{\pi}\frac{\mathrm{d}k}{2\pi}\,\big[(\epsilon_k^\tau)^2-2\epsilon_k^\tau \epsilon_k^0\cos\Delta_k+(\epsilon_k^0)^2\big],
\end{align}\end{subequations}
showing that, to leading order,  all quantities  scale with the same order in $\beta$. 
Note also that these expressions do not assume the quench is infinitesimal; only that it is instantaneous. 
Next, let us specialize to the case of an infinitesimal quench in $g$. 
That is, we set  $g_{\tau}=g_0+\delta g$, $\delta g\ll1$ and $\gamma_\tau = \gamma_0$.
In this case we get  $\sin\Delta_k \simeq -\delta g\,\gamma_0 \sin k/(\epsilon_k^0)^2$ so that Eqs.~\eqref{CSb}-\eqref{IrrSb} simplify to 
\begin{subequations}
\begin{align}
 &\mathcal{C}(\rho')=\beta^2\delta g^2\int_0^{\pi}\frac{\mathrm{d}k}{2\pi}\,\gamma_0^2\frac{\sin^2k}{(\epsilon_k^0)^2},\label{CohSbSg}\\[1em]
 &\mathcal{D}(\rho')=\beta^2\delta g^2\int_0^{\pi}\frac{\mathrm{d}k}{2\pi}\,\frac{(g_0-\cos k)^2}{(\epsilon_k^0)^2},\\[1em]
 &\Delta\mathcal{S}_{\text{irr}}=\frac{1}{2}\beta^2\delta g^2.
 \label{IrrSbSg}
\end{align}
\end{subequations}
From \eqref{CohSbSg} it is clear that for this type of quench, the coherence term is maximal for the Ising model ($\gamma_0=1$), decreasing monotonically with $\gamma_0$ until it vanishes in the XX case ($\gamma_0=0$). 
In particular, for $\gamma_0= 1$, the integral in Eq.~\eqref{CohSbSg} may be evaluated analytically, to give
\begin{equation}\label{CohLimHalf}
 \mathcal{C}(\rho')|_{\gamma_0=1}
 = \begin{cases}
 \frac{\beta^2\delta g^2}{4} &\text{for }|g_0|\leq1, \\[0.2cm]
\frac{\beta^2\delta g^2}{4g_0^2}  & \text{for }|g_0|>1.
 \end{cases}
\end{equation}
This result is quite interesting. 
First, comparing with Eq.~(\ref{IrrSbSg}), we see that when $|g_0| \leq 1$,  half of all the non-equilibrium lag is due to quantum coherence. 
This is somewhat counterintuitive since this is the high-temperature limit, where one would expect quantum coherent effects to play a marginal role. 

\begin{figure}[h!]
 \centering
 \subfigimg[width=0.45\linewidth]{(a)}{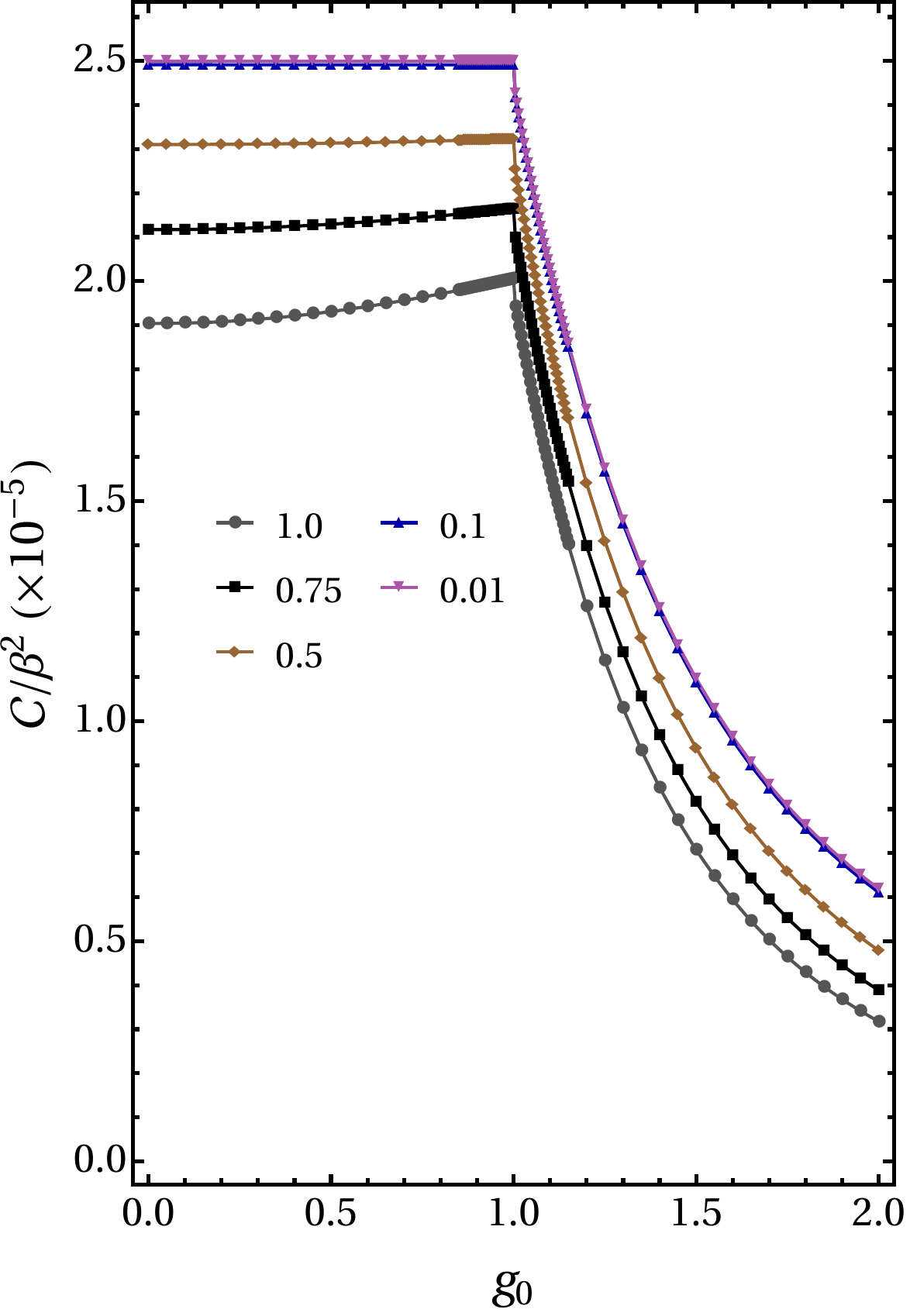}%\\
 \subfigimg[width=0.45\linewidth]{(b)}{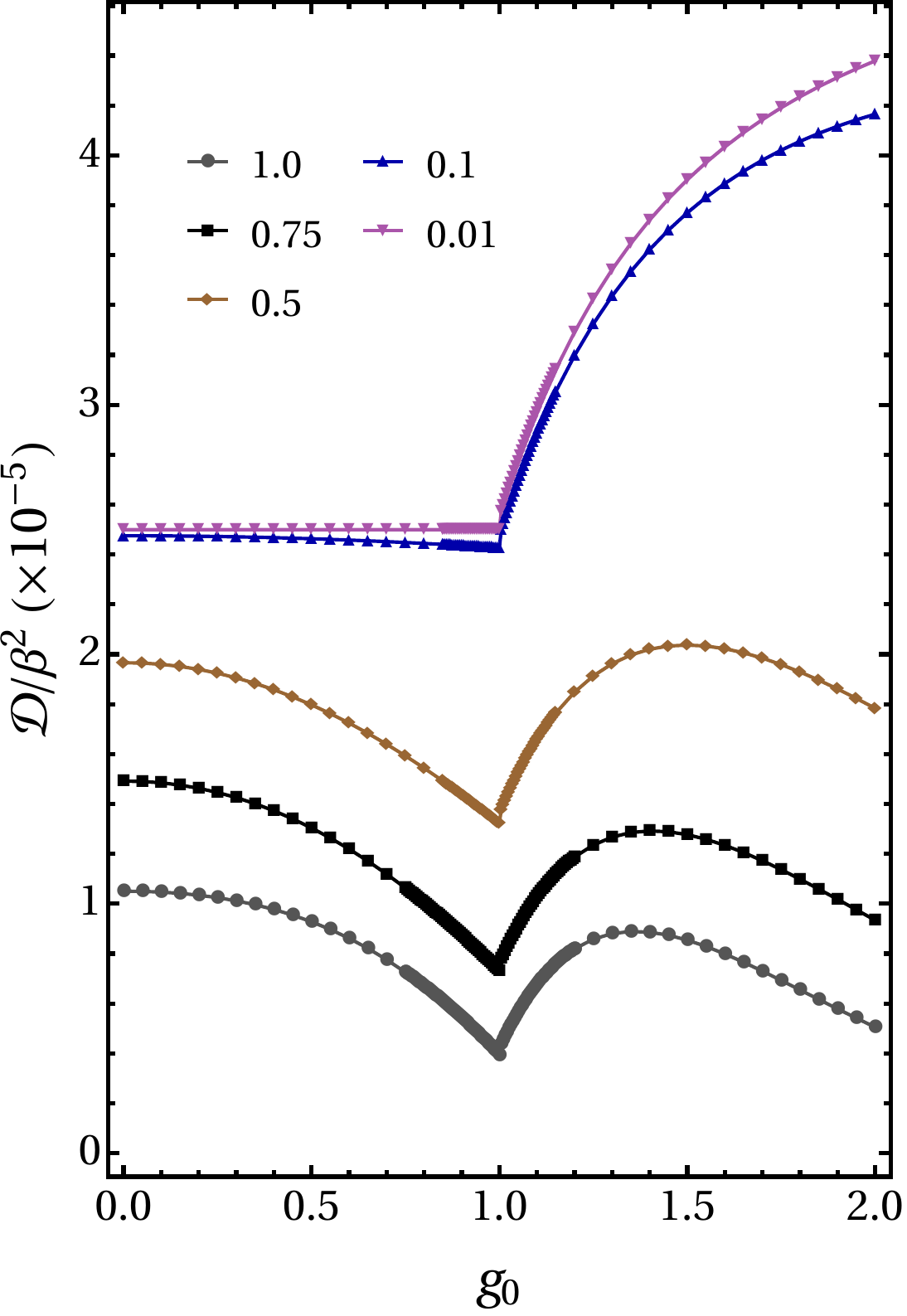}
 %\subfigimg[]{(a)}{\label{cohesmallbeta}\includegraphics[keepaspectratio, width=8cm]{CSmall.pdf}}\\
 %\subfigimg[]{(b)}{\label{difpopsmallbeta}\includegraphics[keepaspectratio, width=8cm]{DSmall.pdf}}
\caption{\label{C&DSmall} High temperature behavior of (a) $\mathcal{C}(\rho')/\beta^2$ and (b) $\mathcal{D}(\rho')/\beta^2$ as functions of $g_0$, computed using Eqs.~\eqref{GenCoh} and \eqref{GenPop} for quenches in $g$ with amplitude $|g_{\tau}-g_0|=0.01$  and fixed $\gamma_0 = 1$.
Different curves correspond to different values of $\beta$. 
The curves in (a)  approach Eq.~\eqref{CohLimHalf} for high temperatures. 
Both quantities present a kink at the critical point.
}
\end{figure}

Second, and perhaps even more impressive, we see that Eq.~(\ref{CohLimHalf}) behaves differently in the two phases. 
And while being continuous, it presents a kink at the critical point. 
This behavior is plotted in Fig. \ref{C&DSmall}(a). 
Results for $\mathcal{D}(\rho')/\beta^2$ in the same range of parameters are  presented in Fig. \ref{C&DSmall}(b).
The high-temperature behavior of the coherence term therefore reflects the nature of the quantum phase transition (which occurs at zero temperature). 
We are unable to provide an intuitive justification for this behavior. 
And to the best of our knowledge, we are unaware of any other high temperature quantities which present non-analyticities at a quantum critical point.
Of course, whether this behavior is experimentally assessable is a complicated question, which has to be addressed in a case-by-case basis. 
In general $\mathcal{C}(\rho')$ is not directly related to an observable, so that measuring it experimentally will in general be highly non-trivial (requiring full state tomography). 
However, $\mathcal{D}(\rho')$ also presents similar signatures and, in principle, is much more easily measurable since it depends only on measurements in the energy basis.

We can similarly perform a quench in the anisotropy parameter, keeping $g_\tau = g_0$ and setting $\gamma_\tau = \gamma_0 + \delta \gamma$. In this case we get $\sin\Delta_k \simeq \delta\gamma(g_0-\cos k)\sin k/(\epsilon_k^0)^2$. 
Eqs.~(\ref{CSb})-(\ref{IrrSb}) then simplify to 
\begin{subequations}\label{SmallBetaSmallAni}
 \begin{align}
  &\mathcal{C}(\rho')=\beta^2\delta\gamma^2\int_0^{\pi}\frac{\mathrm{d}k}{2\pi}\,\frac{(g_0-\cos k)^2\sin^2k}{(\epsilon_k^0)^2},\\[1em]
  &\mathcal{D}(\rho')=\beta^2\delta\gamma^2 \int_0^{\pi}\frac{\mathrm{d}k}{2\pi}\,\gamma_0^2\frac{\sin^4 k}{(\epsilon_k^0)^2},\\[1em]
  &\Delta\mathcal{S}_{\text{irr}}=\frac{1}{4}\beta^2\delta\gamma^2.
 \end{align}
\end{subequations}
What is interesting to note in this case is that if we initially have an XX chain, $\gamma_0=0$, the population mismatch due to the small quench in the anisotropy parameter vanishes, $D(\rho')|_{\gamma_0=0}=0$, and all entropy production is due to coherence, independently of the value of the applied field $g_0$.

The above results show that there is an interplay between $\mathcal{C}$ and $\mathcal{D}$ for high temperatures, as we go from the XX to the Ising model and as we change from a quench in the field to a quench in the anisotropy. For a quench in the field, the coherence contribution to the entropy production vanishes in an XX chain and increases as we go up to the Ising model, where it reaches a maximum, contributing to half the total production of entropy. For a quench in the anisotropy, in contrast, it is $\mathcal{D}$ that vanishes in a initial XX chain, with  all entropy production becoming a consequence of the generation of coherence in the quench protocol. As $\gamma_0$ is increased, $\mathcal{C}$ steadily decreases, reaching a minimum for the Ising model.

%nova subseção
%nova subseção
%nova subseção
%nova subseção

\subsection{\label{LargeBeta}Low temperature limit}
For large $\beta$, Eqs.~\eqref{GenCoh}-\eqref{GenPop} can be approximated by 
\begin{subequations}\begin{align}
  &\mathcal{C(\rho')}=
  %\frac{1}{\pi}\int_0^{\pi}\mathrm{d}k\,\mathcal{C}_k(\rho')=
  \int_0^{\pi}\frac{\mathrm{d}k}{2\pi}\,[-p_k\ln p_k-(1-p_ k)\ln(1-p_k)],\label{LargeBCoh}\\[1em]
  &\mathcal{D(\rho')}=%\frac{1}{\pi}\int_0^{\pi}\mathrm{d}k\,\mathcal{D}_k(\rho')=
  4\beta\int_0^{\pi}\frac{\mathrm{d}k}{2\pi}\,\epsilon_k^\tau p_k-\mathcal{C}(\rho'),\\[1em]
  &\Delta\mathcal{S}_{\text{irr}}=4\beta\int_0^{\pi}\frac{\mathrm{d}k}{2\pi}\,\epsilon_k^\tau p_k\label{EntProdLargeBeta}.
\end{align}\end{subequations}
where $p_k=\sin^2\big(\Delta_k/2\big)$. 
Quite interestingly, the integrand in Eq.~\eqref{LargeBCoh} is seen to be nothing but the binary Shannon entropy  associated with the two-point distribution $(p_k,1-p_k)$ (for each $k$).
The physical interpretation of $p_k$ can be understood from Eq.~\eqref{pre_post_states}, which shows that  $p_k= \sin^2\big(\Delta_k/2\big)$ is nothing but the probability of the unoccupied (occupied) pre-quench modes $\pm k$ to become occupied (unoccupied) after the quench. 
With this picture in mind, the non-equilibrium lag~(\ref{EntProdLargeBeta}) is seen to result solely  from this change in occupation, whereas the coherence reflects the entropy associated with this  occupation probability. 

A notable thing about Eq. \eqref{LargeBCoh}, is that it does not depend on $\beta$, unlike $\mathcal{D}(\rho')$ and $\Delta\mathcal{S}_{\text{irr}}$. This means that, as the  temperature is decreased, the relative contribution of $\mathcal{C}(\rho')$ to  $\Delta\mathcal{S}_{\text{irr}}$ becomes increasingly less important. 
%This happens because, at $T=0\,(\beta\rightarrow\infty)$, the system would have to be in its ground state. Then, any change in the Hamiltonian would necessarily lead to excitations, which in turn would mean a great contribution  from the population difference $\mathcal{D}$ and, therefore, a limited contribution from $\mathcal{C}$.

\begin{figure}
 \centering
 \subfigimg[width=0.45\linewidth]{(a)}{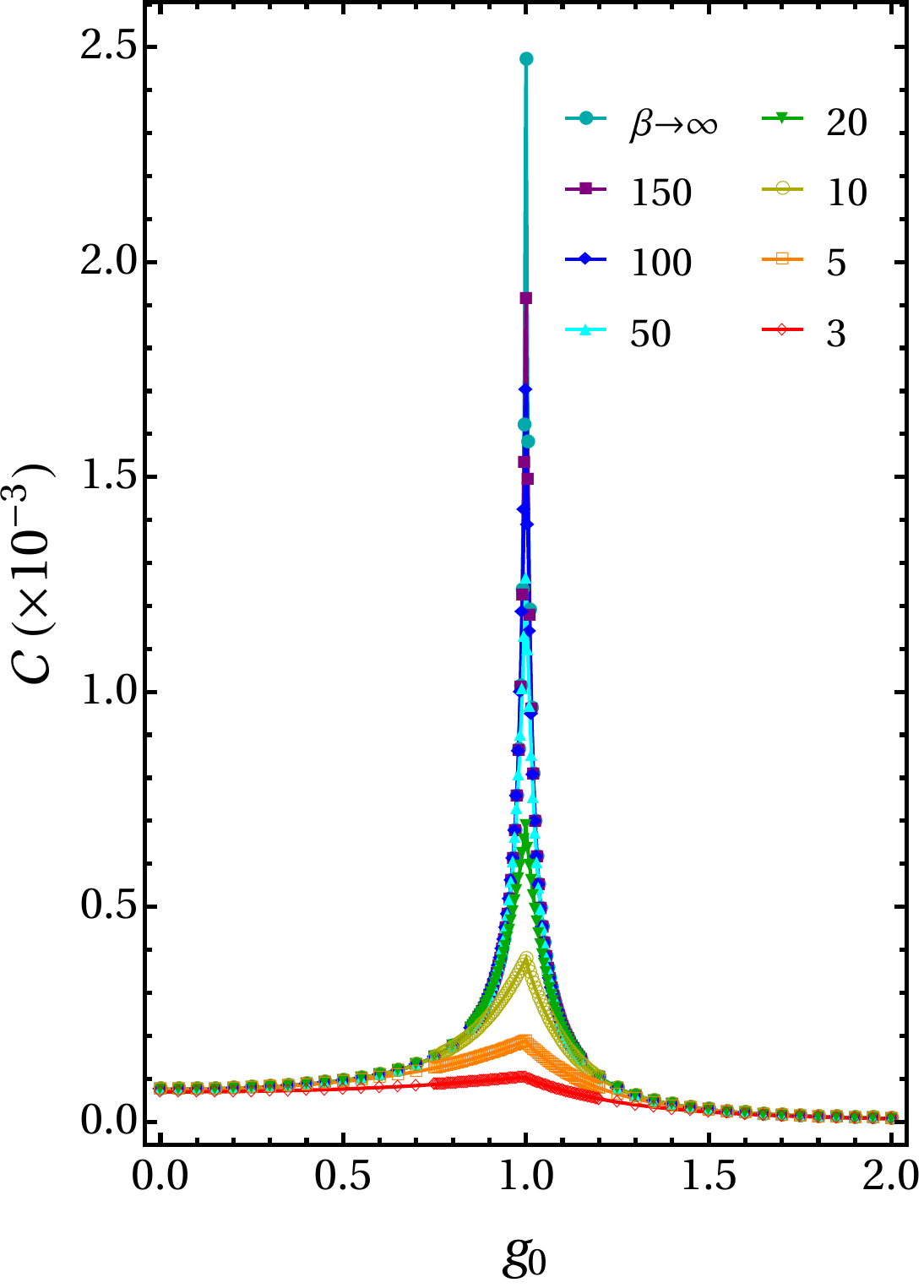}%\\
 \subfigimg[width=0.45\linewidth]{(b)}{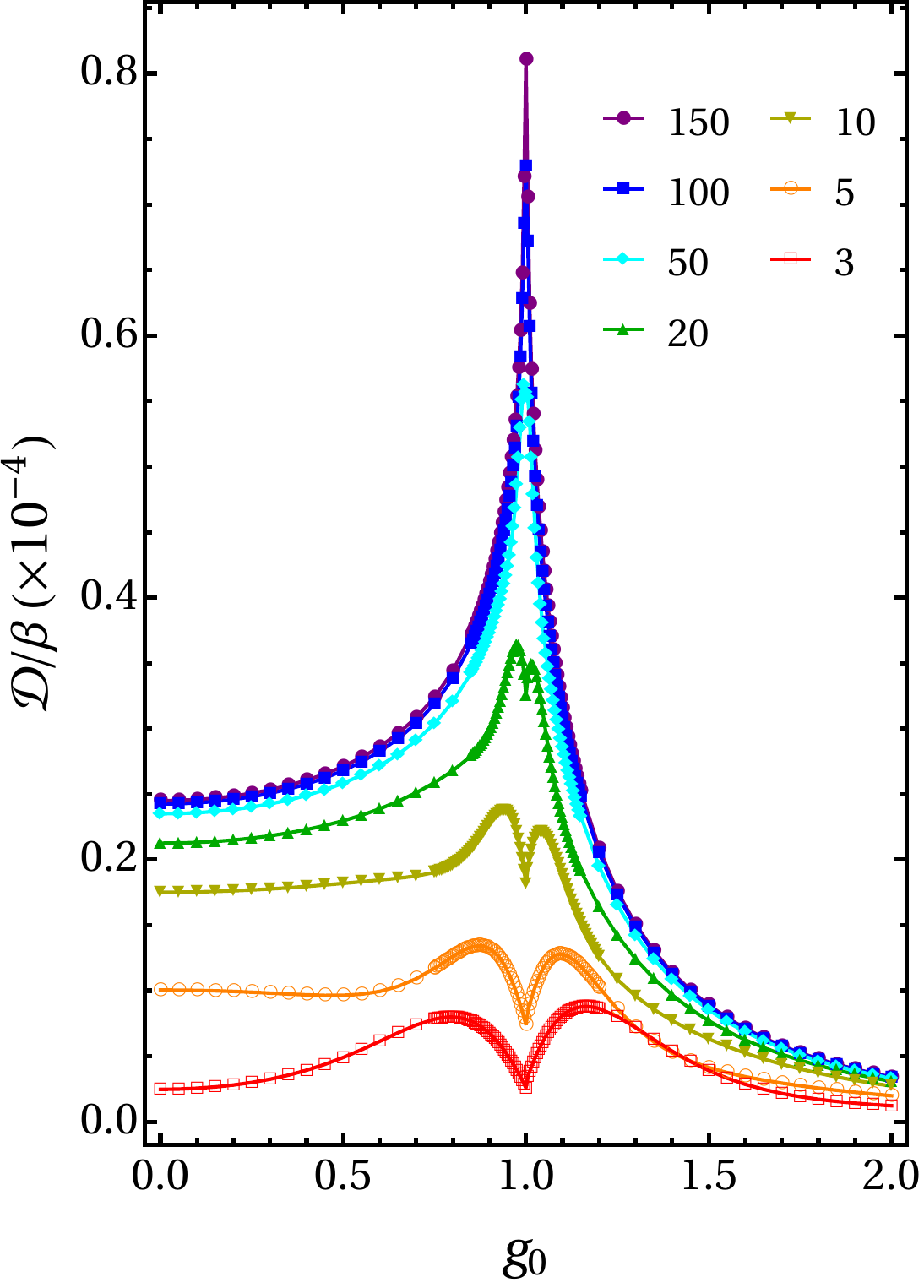}
 %\subfloat[ $\mathcal{C}(\rho')$ for the Ising model.]{\includegraphics[keepaspectratio, width=8cm]{CLarge.pdf}}\\
 %\subfloat[]{\includegraphics[keepaspectratio, width=8cm]{DLarge.pdf}}
\caption{\label{C&DLarge}Low temperature behavior of (a) $\mathcal{C}(\rho')$ and (b) $\mathcal{D}(\rho')/\beta$ for several $\beta$ as functions of $g_0$, computed using Eqs.~\eqref{GenCoh} and \eqref{GenPop} for quenches in $g$ with amplitude $|g_{\tau}-g_0|=0.01$  and fixed $\gamma_0 = 1$.
The curves in (a) approach Eq.~\eqref{LargeBCoh} in the limit  $T=\frac{1}{\beta}\rightarrow0$. 
} 
\end{figure}

We start our analysis of Eqs.~(\ref{LargeBCoh})-(\ref{EntProdLargeBeta}) by considering quenches in $g$, with $\gamma_0 = 1$ (Ising). 
The results are shown in Fig.~\ref{C&DLarge}, where we plot $\mathcal{C}(\rho')$ and $\mathcal{D}(\rho')/\beta$. 
Clearly, as $\beta \to \infty$ the latter becomes dominant. 
As a consequence $\Delta\mathcal{S}_{\text{irr}} \simeq \mathcal{D}(\rho')$.
This is a consequence of the fact that, in this case, changes in the Hamiltonian lead to a significant production of excitations, thus causing the contribution from populations to become dominant. 
Indeed, in this limit the non-equilibrium lag is known to be proportional to the magnetic susceptibility $\chi = - \partial^2 F/\partial g_0^2$  (where $F$ is the equilibrium free energy), according to the relation    \cite{Gambassi1106,PhysRevX4031029}
\begin{equation}
\Delta\mathcal{S}_{\text{irr}}=\beta\delta g^2\chi.
\end{equation}
As a consequence, $\Delta\mathcal{S}_{\text{irr}}/\beta\delta g^2$ diverges logarithmically around the critical points $|g_0|=1$ \cite{Pagnelli,PhysRevE92022108,Mascarenhas1307}. 
This divergence is due solely to the changes in populations. 

\begin{figure}[h!]
 \centering
 \subfigimg[width=0.45\linewidth]{(a)}{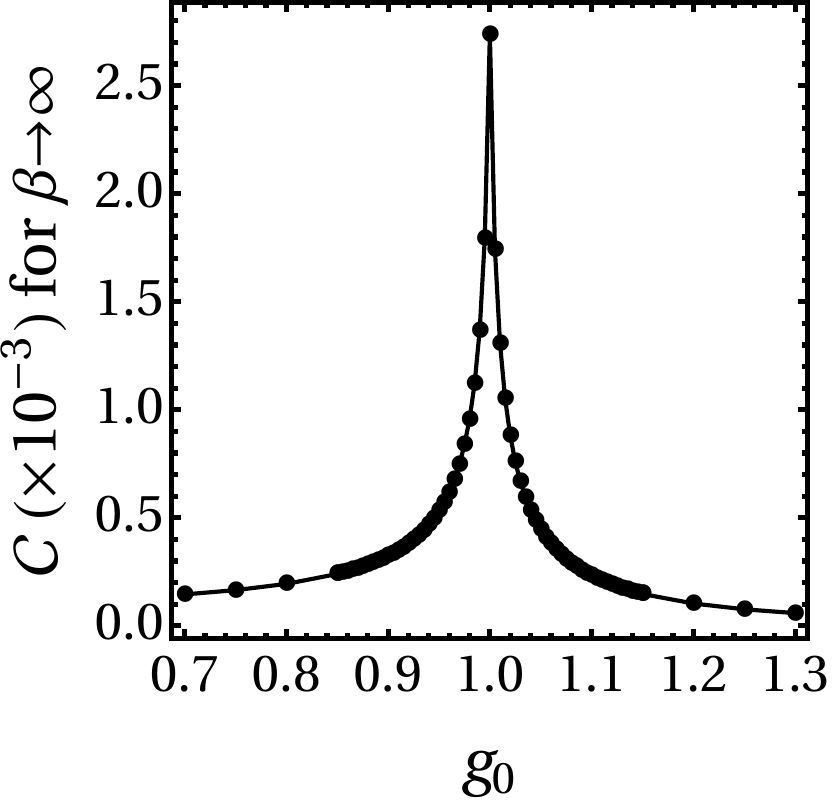}
 \subfigimg[width=0.45\linewidth]{(b)}{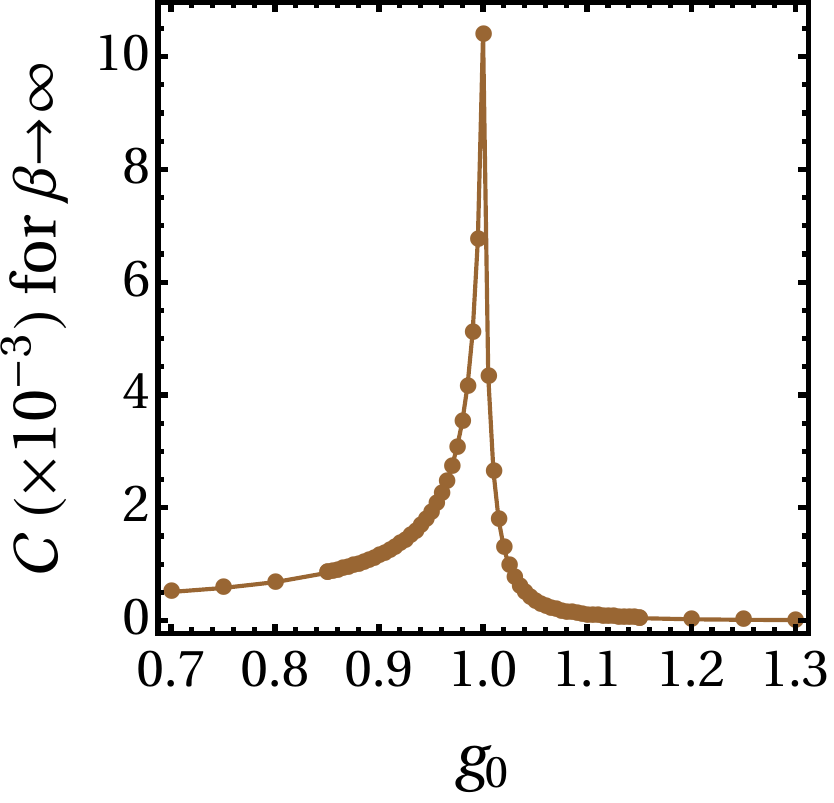}
\caption{\label{CohLargeXX}$\mathcal{C}(\rho')$~vs.~$g_0$ for $T = 0$ and quenches in $g$ of magnitude $|g_{\tau}-g_0|=0.01$. 
(a) Close to the Ising case, $\gamma_0 = 0.9$ and (b) close to the XX case, $\gamma = 0.2$. 
The curves show how the cusp of $\mathcal{C}(\rho')$ becomes more asymmetric as $\gamma_0$ is reduced.} 
\end{figure}

The coherence in Fig.~\ref{C&DLarge}(a), on the other hand, does not diverge, which we emphasize by including a plot of $\beta \to \infty$ in Fig.~\ref{C&DLarge}(a). 
Instead, $\mathcal{C}(\rho')$ shows a cusp at the critical point.
In fact, Eq.~\eqref{LargeBCoh} is bounded from above by $\frac{1}{2}\ln 2$, with this maximum value occurring only for $p_k=1/2$ for all $k$'s.
From our numerical analysis we also find that the height of the cusp at $g_0 = 1$ scales linearly with $\delta g$.  
The shape of the cusp in $\mathcal{C}(\rho')$ depends on the value of $\gamma_0$.  This is presented in Fig. \ref{CohLargeXX}, where we plot $\mathcal{C}(\rho')$ for $\beta\to\infty$ for different values of $\gamma_0$.
As can be seen, it changes from a very symmetric form for larger values of $\gamma_0$ to an increasingly asymmetric format as $\gamma_0$ decreases.
%The existence of this cusp, in fact, depends on the value of $\gamma_0$ and may actually change to a discontinuity. 
%This is presented in Fig. \ref{CohLargeXX}(a), where we plot $\mathcal{C}(\rho')$ for $\beta \to \infty$ (still for infinitesimal quenches in $g$), but with varying $\gamma_0$. 
%The cusp remains for any finite $\gamma_0$, but asymptotically becomes a discontinuity in the limit $\gamma_0 \to 0$. 
%{\color{red}[GTL: do we really need fig 4(b)?]}.

\begin{figure}[h!]
 \centering
 \subfigimg[width=0.45\linewidth]{(a)}{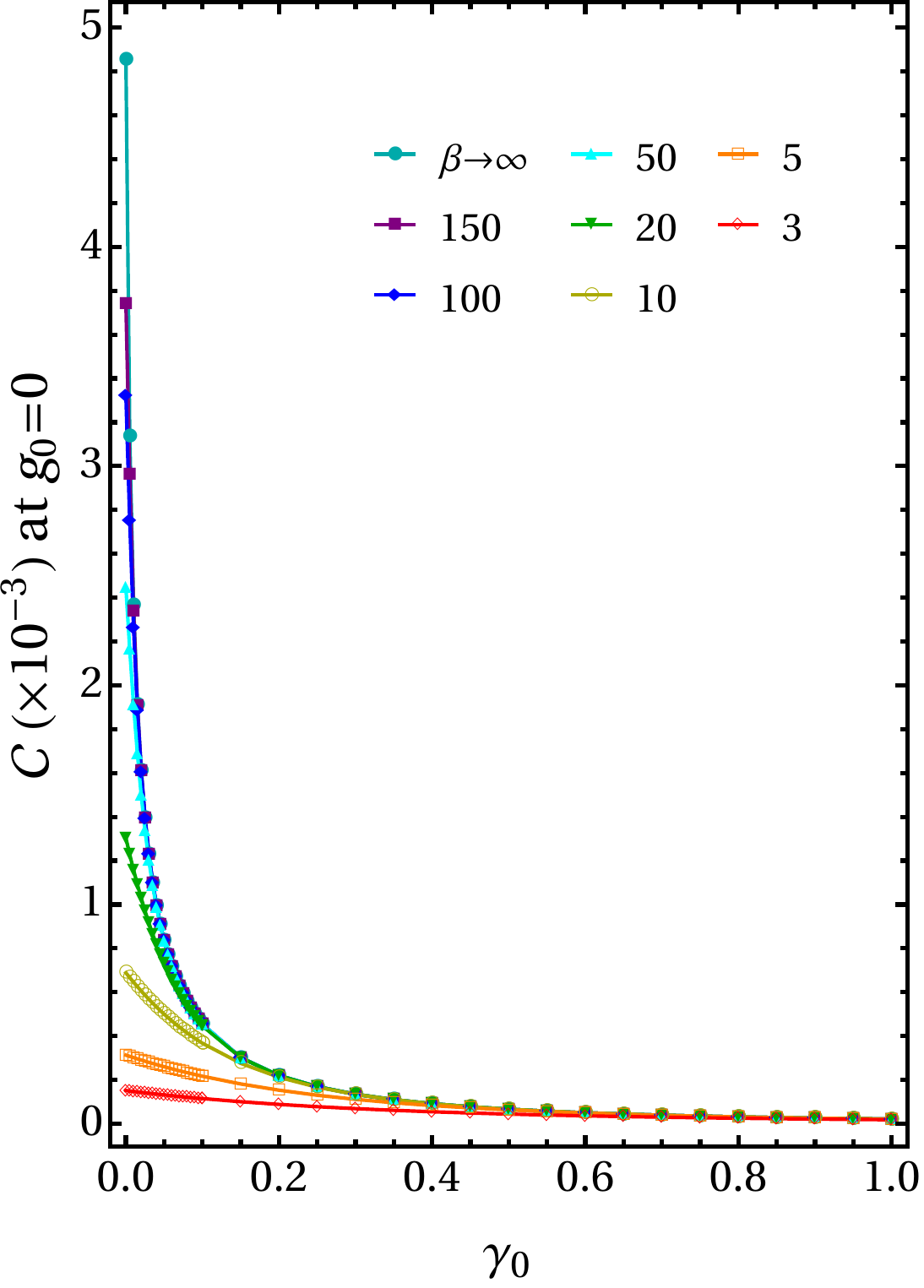}
 \subfigimg[width=0.45\linewidth]{(b)}{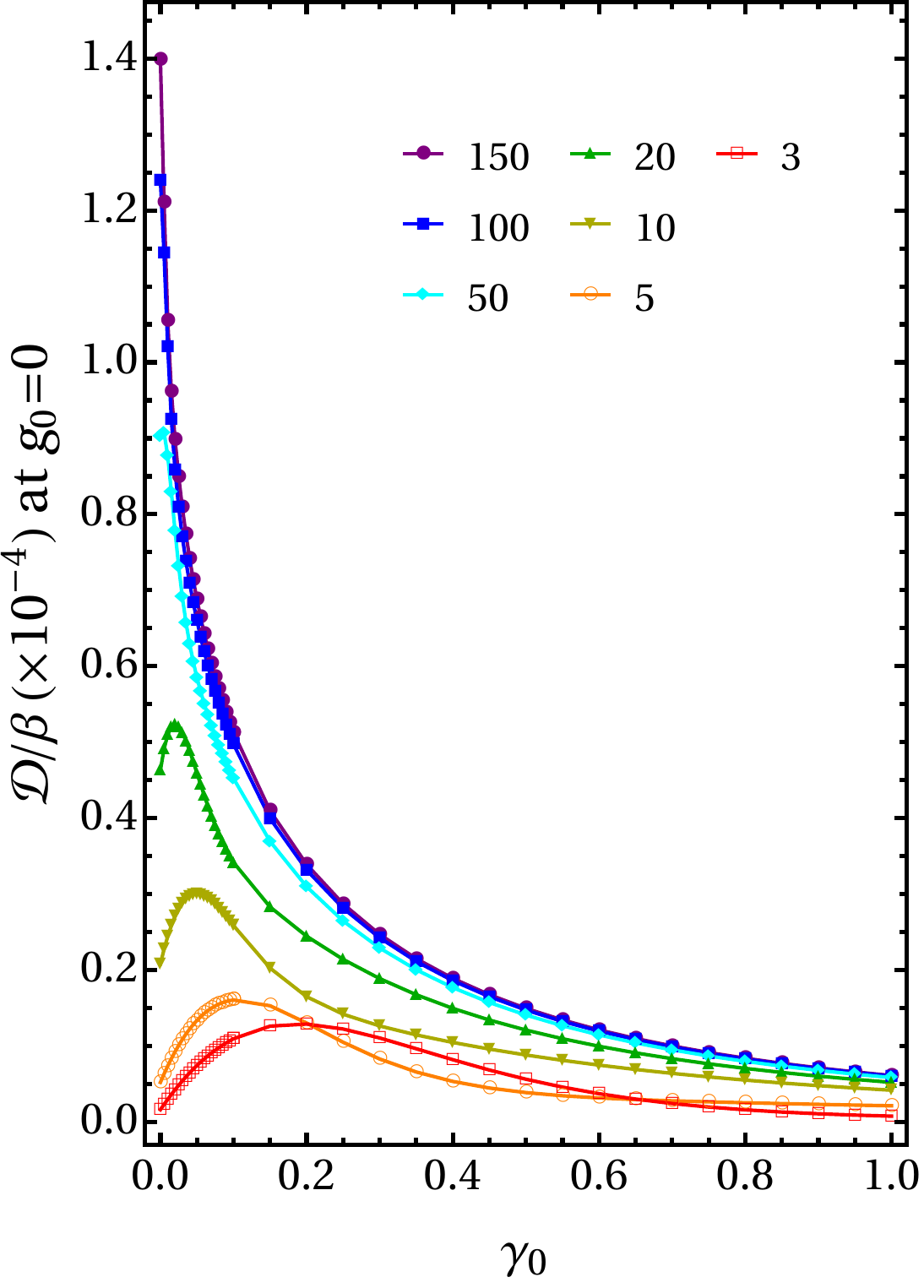}
\caption{\label{C&DLargeAni}
Low temperature behavior of (a) $\mathcal{C}(\rho')$ and (b) $\mathcal{D}(\rho')/\beta$ for several $\beta$ as functions of $\gamma_0$, computed using Eqs.~\eqref{GenCoh} and \eqref{GenPop} for quenches in $\gamma$ with amplitude $|\gamma_{\tau}-\gamma_0|=0.01$  and fixed $g_0 =0$.
} 
\end{figure}

We also studied the case of quenches in the anisotropy, with fixed field. In this context, the coherence decreases with increasing $\gamma_0$ and has its maximal value for a vanishing field, see Fig. \ref{C&DLargeAni}.

%nova subseção
%nova subseção
%nova subseção
%nova subseção

\subsection{\label{RatioCS}Ratio $\mathcal{C}(\rho')/\Delta\mathcal{S}_{\text{irr}}$}

\begin{figure}
\centering
 \includegraphics[keepaspectratio, width=0.9\linewidth]{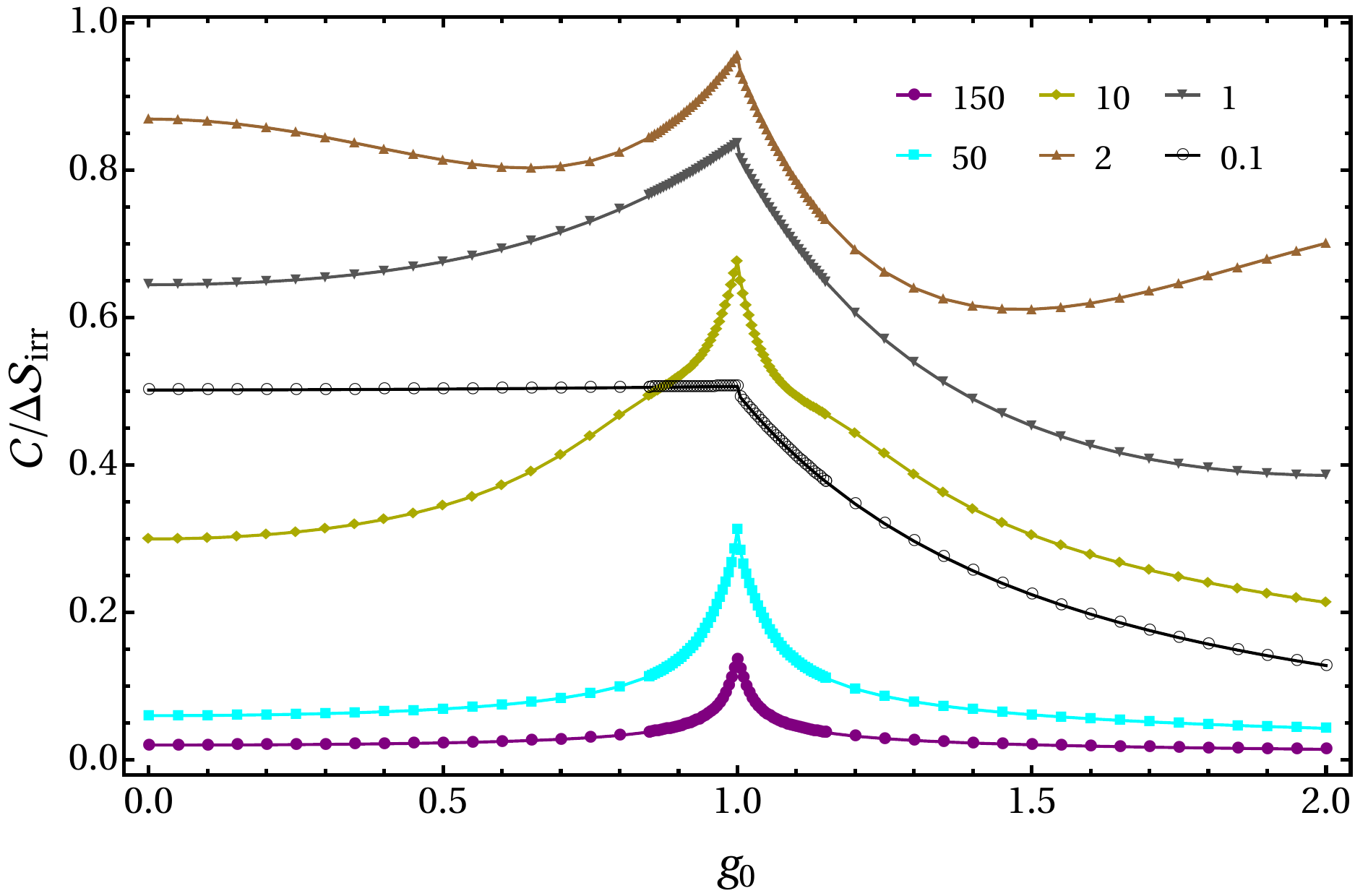}
 \caption{\label{CSIsing}Ratio $\mathcal{C}/\Delta S_\text{irr}$~vs.~$g_0$ for several $\beta$ and quenches in $g$ with magnitude $|g_{\tau}-g_0|=0.01$, with fixed $\gamma_0 = 1$.
For  $\beta=0.1$ the ratio approaches the limit in Eq. \eqref{CohLimHalf}. A notable feature is the presence of an optimal temperature $\beta\approx2$, in which almost all entropy production is due to the generation of coherence.}
\end{figure}

Next we combine the high and low temperature results and perform an analysis of the relative contribution $\mathcal{C}(\rho')/\Delta\mathcal{S}_{\text{irr}}$.  
Results for quenches in $g$, with $\gamma_0 = 1$ and several values of $\beta$ are shown in Fig.~\ref{CSIsing}. 
In the case of high temperatures, e.g. $\beta=0.1$, this fraction approaches $1/2$, which is the limit predicted by Eq. \eqref{CohLimHalf}. 
Similarly, for low temperatures, the ratio tends to zero, as discussed in Sec.~\ref{LargeBeta}. 
The notable features of Fig.~\ref{CSIsing}, however, is for intermediate temperatures, where the results are not at all intuitive. 
First, there exists an ``optimal'' temperature, around $\beta\approx2$, for which the ratio approaches unity, so that almost all entropy produced stems from coherence. 
This happens because as the dependence of $\mathcal{D}(\rho')$ on $\beta$ changes from $\beta^2$, in the high temperature limit, to a linear dependence on $\beta$ in the low temperature limit, there is a range of temperatures in which the coherence generation for a given quench increases more rapidly than the population imbalance. Second,  for large $\beta$, even though the ratio is  generally small, there is nonetheless a substantial increase in the vicinity of the critical point. 
This is a consequence of the sharp peak in the coherence in this region, as shown in Fig. \ref{C&DLarge}. 
However, since the coherence saturates for increasing $\beta$ while the entropy production always increases, this peak in the fraction $\mathcal{C}/\Delta\mathcal{S}_{\text{irr}}$ approaches zero as the temperature tends to zero.

\begin{figure}
\centering
 \includegraphics[keepaspectratio, width=0.9\linewidth]{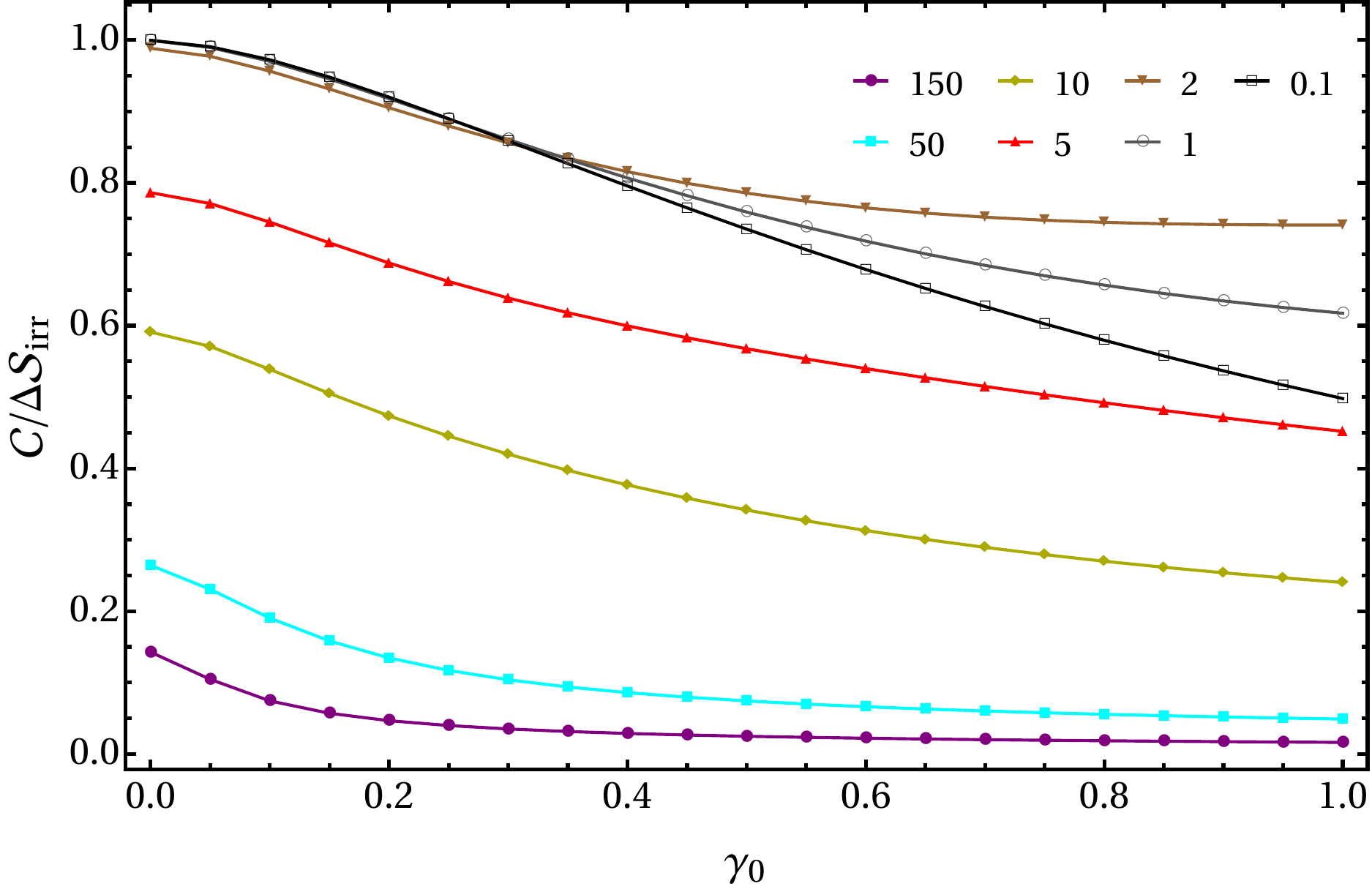}
 \caption{\label{CSAni}$\mathcal{C}(\rho')/\Delta\mathcal{S}_{\text{irr}}$ as a function of $\gamma_0$ for a series of quenches in the anisotropy parameter with amplitude $|\gamma_{\tau}-\gamma_0|=0.01$, for several $\beta$, at $g_0=1$. It shows that, for small $\beta$ and $\gamma_0=0$, all entropy production is due to the generation of coherence for this type of quench.
 }
\end{figure}

A similar analysis for quenches in the anisotropy parameter is shown in Fig.~\ref{CSAni}.
% The curves for $\beta=0.1$ and $\beta=0.01$ show how the ratio approaches unity as $\gamma_0 \to 0$, as previously discussed in Sec.~\ref{SmallBeta}.
The curve for $\beta=0.1$ show how the ratio approaches unity as $\gamma_0 \to 0$, as previously discussed in Sec.~\ref{SmallBeta}.
Notably, for intermediate values of $\beta$, between $\beta=5$ to $\beta=10$, in the critical point, the coherence accounts for a large part of the production of entropy, between $25\%$ to $80\%$, for any value of the initial anisotropy. Again for large $\beta$, this ratio approaches zero.

% Figure \ref{CSAni2} illustrates the change in the ratio $\mathcal{C}(\rho')/\Delta\mathcal{S}_{\text{irr}}$ for quenches with $|\gamma_{\tau}-\gamma_0|=0.01$ for several pairs $(\beta,\,\gamma_0)$ as a function of the field $g_0$. It shows that for small and large $\beta$ - $\beta=0.1$ and $\beta=100$, respectively, this fraction has its maximal value in the vicinity of the critical point $g_0=1$. It also shows that for this type of quench, the coherence generation is optimized for smaller anisotropy parameter $\gamma_0$, which is consistent with the results in Eq. \ref{SmallBetaSmallAni}. That is, for quenches in the anisotropy, the generation of coherence is larger if we initially have an XX chain  and becomes smaller as we increase $\gamma_0$.
% \begin{figure}[h!]
% \centering
%  \includegraphics[keepaspectratio, width=8cm]{RatioCSAni2.pdf}
%  \caption{\label{CSAni2}The fraction of coherence generation as a function of the applied field $g_0$ for a series of quenches in the anisotropy with amplitude $|\gamma_{\tau}-\gamma_0|=0.01$ for several pairs $(\beta,\,\gamma_0)$. It shows how, for small $\beta$, this fraction approaches $1$ for any value of the transverse field.}
% \end{figure}
%nova seção
%nova seção
%nova seção
%nova seção
%nova seção
%nova seção

\section{\label{Conc}Conclusion}
We investigated the genuinely quantum-mechanical contribution of the generation of coherence to the production of entropy for quenches in the transverse field and in the anisotropy parameter of an $XY$ model. We showed that the generation of coherence is intimately related to the rotation in the basis that diagonalizes the system's Hamiltonian when the quench protocol is performed.

For large temperatures (small $\beta=1/T$), we showed that there is an interplay between the coherent and incoherent contributions. For small quenches in the transverse field, the coherence increases steadily with the anisotropy parameter, reaching a maximum for the Ising model. For small quenches in the anisotropy, instead, we found that the coherence is the sole responsible for the entropy production if the systems starts in an XX chain. As the initial anisotropy is increased, the coherence decreases and reaches a minimum in the Ising model.

For small temperatures,  we found  a saturation in the contribution from coherence. This results from the fact that in such cases any change in the Hamiltonian leads to excitations on the system, which forcibly makes the production of entropy to be associated with the changes in population on the system. We also showed that the behavior of the coherence around the critical point, for quenches in the field, does not present a discontinuity, but rather a cusp. Notwithstanding, the entropy production still diverges, which is solely due to the changes in populations. 

Finally, we analyzed the relative contribution of coherence to the total entropy production. For quenches in the transverse field in the Ising model, we showed that for small $\beta$ this fraction approaches $1/2$ in the ferromagnetic region. We also found that at certain temperatures  the coherence can account for almost all the entropy production. For quenches in the anisotropy, the ratio of coherence to the production of entropy remains large even for intermediate $\beta$, for any initial anisotropy.

\section*{Acknowledgements}
The authors acknowledge fruitful discussions with M. Perarnau-Llobet, M. Scandi, D. Uip, S. Campbell, L. H. Mandetta and J. Goold. 
We acknowledge financial support from the Brazilian agencies Conselho Nacional de Desenvolvimento Cient\'ifico e Tecnol\'ogico and Coordena\c c\~ao de Aperfei\c coamento de Pessoal de N\'ivel Superior.
GTL and APV acknowledge the S\~ao Paulo Research Foundation FAPESP (grants 2018/12813-0, 2017/50304-7, 2017/07973-5, 2017/07248-9).

% ----------------------------------------------------------
% Referências bibliográficas
% ----------------------------------------------------------
\bibliographystyle{apsrev4-1}
\bibliography{library,bibl}
%\bibliography{library}
\end{document}